# Teaching with AI: A Systematic Review of Chatbots, Generative Tools, and Tutoring Systems in Programming Education

**Said Elnaffar**
Canadian University of Dubai
Dubai, United Arab Emirates

**Farzad Rashidi**
Université Paris Cité
F-75013 Paris, France

**Abedallah Zaid Abualkishik**
American University in the Emirates
Dubai, United Arab Emirates

**Abstract.** This review examines the role of artificial intelligence (AI) agents in programming education, focusing on how these tools are being integrated into educational practice and their impact on student learning outcomes. An analysis of 58 peer-reviewed studies published between 2022 and 2025 identified three primary categories of AI agents: chatbots, generative AI (GenAI), and intelligent tutoring systems (ITS), with GenAI being the most frequently studied. The studies report that the primary instructional objectives include providing enhanced programming support 94.83% of studies, delivering motivational and emotional benefits 18.96%, and increasing efficiency for educators 6.90%. Reported benefits include personalized feedback, improved learning outcomes, and time savings. The review also highlights challenges, such as implementation barriers documented in 93.10% of studies, overreliance resulting in superficial learning in 65.52%, and concerns regarding AI errors and academic integrity. These findings suggest the need for instructional frameworks that prioritize the development of prompt engineering skills and human oversight to address these issues. This review provides educators and curriculum designers with an evidence-based foundation for the practical and ethical integration of AI in programming education.

*Corresponding author: Said Elnaffar; *said.elnaffar@cud.ac.ae*

## 1. Introduction

Several factors in programming education create a disconnect between learning goals and student achievement. Programming education faces ongoing challenges. Students frequently struggle to understand basic concepts despite the implementation of various teaching approaches. Notably, dropout rates in introductory programming courses can reach 30 to 50 percent. This high attrition rate raises questions about the efficacy of existing pedagogical approaches in supporting learners. Consequently, there is a need to implement innovative strategies to enhance student engagement and facilitate the acquisition of programming concepts (Watson et al., 2014).

The primary limitations often cited include students' limited exposure to computational thinking and structural deficiencies in educational infrastructure. These factors create confusion and slow down learning progress. Furthermore, it is difficult for current assessment tools to gauge how these two challenges interact: students' unfamiliarity with programming concepts and their limited resources. Additionally, many students adopt a fixed mindset when facing complex algorithmic problems, which further impede their learning. Concurrent processing of multiple cognitive demands can exceed working memory capacity, resulting in cognitive overload. This increased workload may reduce academic performance and increase mental fatigue, which in turn can lead to student disengagement and eventual withdrawal from computer-related disciplines (Qian & Lehman, 2017).

Insufficient funding in many educational institutions intensifies educational disparities among socio-economic groups. Such institutions often face shortages of qualified teachers, inadequate facilities, and outdated educational resources, collectively undermining the quality of instruction and student learning. Recent research and case studies (Bakar et al., 2025; Nurmamatov et al., 2025) corroborate these challenges (Bakar et al., 2025; Hudin, 2023; Xayriddin, 2025).

AI can transform the policies and methods of programming education. Tools like ChatGPT, Copilot, and other AI systems can help address challenges that traditional teaching methods have struggled to overcome. They provide instant feedback, handle repetitive coding tasks, and assist students in solving problems. This support is particularly important for students with learning disabilities. New tools can compensate for the shortcomings of traditional teaching, but they may also introduce new challenges; therefore, more research and investigation are needed to utilize these tools effectively (Chang et al., 2024; Stray et al., 2025; Zhai, 2023).

AI is increasingly influential in coding education, yet its adoption raises several critical questions. The technology is anticipated to reduce cognitive load and foster greater student engagement. Preliminary evidence suggests that integrating AI can enhance learning and academic performance (Thomas et al., 2024). However, the optimal strategies for implementing these tools in programming courses remain uncertain. Insufficient data makes it difficult to draw definitive conclusions regarding the alignment of AI technologies with educational objectives or their effectiveness across diverse educational contexts. Additionally, deployment in under-resourced settings may introduce unique challenges. As a result, targeted research is necessary to inform context-specific implementation.

These central questions motivate the present literature review, which aims to elucidate the role of AI agents in programming education. While general reviews of AI in education often overlook the unique challenges associated with programming instruction, this study explicitly examines conversational agents, generative assistants, and ITS. Beyond cataloguing applications, the review articulates their educational objectives, delineates practical advantages and limitations, and highlights their substantial influence on instructional practices and disciplinary knowledge in this evolving domain.

## 2. Review of Literature

This review critically examines research on programming education, highlighting both advancements and ongoing challenges. It discusses the evolution of research addressing persistent instructional issues and analyzes the emergence of digital tools, particularly AI, as potential solutions to these issues. The analysis considers both established findings and enduring gaps in literature. The primary objectives are to acknowledge prior research and to identify key areas requiring further investigation. This review identifies a paucity of focused studies on AI implementation in coding classes, indicating a gap that warrants further research.

### 2.1 Challenges in Programming Education

Introductory programming courses pose significant challenges for both students and instructors. According to Luxton-Reilly et al. (2018), these challenges arise from the inherent complexity of programming, ineffective teaching methodologies, insufficient feedback, and inadequate assessment practices. The review emphasizes the importance of formative feedback, academic integrity, and student-centred teaching methods. Medeiros et al. (2018), after reviewing 89 studies, identified three main categories of challenges in programming education. The first category includes cognitive difficulties, such as learning syntax, logical reasoning, and debugging. The second involves instructional limitations, like a lack of personalized support. Finally, they noted emotional factors, including low motivation and reduced self-confidence. Separately, the integration of computational thinking is also considered essential for improving critical thinking skills.

### 2.2 Digital and Data-Driven Tools in Programming Education

Researchers have proposed integrating digital tools into educational practices to address current challenges. Ihantola et al. (2015) investigated the role of educational data mining and learning analytics in supporting computer science education. Their findings indicate that these technologies can aid educators by providing automated feedback, tracking students' academic progress, and identifying potential issues. Nevertheless, concerns remain regarding data collection and privacy. Collaboration between researchers and practitioners is essential to ensure the ethical and practical implementation of these technologies. Similarly, Asgari et al. (2024) explored programming students' beliefs of digital tools in Sweden and Taiwan. Their study found that immediate feedback for error correction and the demonstration of complex concepts was perceived as the most effective learning methods. Students also identified online discussion platforms and gamification as valuable for fostering collaborative and engaging learning environments. These digital tools facilitate the application of theoretical knowledge in practical contexts. However, concerns persist about potential bias in AI-based grading systems and the need for more precise guidance on the appropriate use of technology

**2.3 AI in Education**
AI has been integrated into various sectors, including education, where it can enhance instructional practices and support learning outcomes. In a study on programming education, Silva et al. (2024) examined undergraduate students' use of ChatGPT to solve coding tasks. Most participants reported higher productivity and less frustration when using the tool. However, the study also identified potential limitations. These include over-reliance on AI-generated solutions and a decline in the growth of independent problem-solving skills. The authors recommended that AI tools serve as helpful learning aids rather than replacements for active learner engagement.

A systematic review by Manorat et al. (2025) analyzed 119 studies (2012-2024) on AI uses in programming education, finding four key areas: first, automated course design using large language models, secondly, classroom personalization through adaptive content recommendation, thirdly, AI-powered assessment and feedback generation, and fourthly, predictive analytics for student performance monitoring. Their findings show that while AI significantly reduces teachers' grading workload (e.g., achieving 74.9% accuracy in automated code assessment), its most transformative impact lies in enabling real-time, scaffolded support for learners through virtual assistants and personal exercise recommendations. The review cautions against uncritical adoption, echoing Silva et al.'s concerns about potential overreliance, and emphasizes the need for teaching frameworks to guide the integration of AI.

In summary, although extensive research exists on the use of AI in education, most studies focus on general classroom applications or subjects such as language learning and mathematics. Few studies have systematically investigated the implications of AI agents in programming education. This review addresses this gap by synthesizing available evidence and providing a structured framework for understanding the role of AI agents in programming education.

To guide this review, the following research questions were formulated:
**RQ1**: What types of AI agents are being used in programming education, and how are they categorized?
**RQ2**: What instructional objectives and pedagogical benefits are reported for the use of AI agents in programming education?
**RQ3**: What implementation challenges and risks are identified in the literature concerning AI agents in programming education?

## 3. Methodology

This study employed a Systematic Literature Review (SLR) based on the Preferred Reporting Items for Systematic Reviews and Meta-Analyses (PRISMA) 2020 guidelines to synthesize existing research on the use of AI agents in programming education (Tugwell & Tovey, 2021). The goal is to identify and categorize the various types of AI tools used in coding classes. This review examines the goals and learning outcomes of the technologies used in programming education. The review also examines the benefits and challenges associated with using AI tools in coding classes. This section outlines the method employed in this review, including the search strategy, selection criteria, and data verification process.

### 3.1 Search Strategy

The literature search was conducted across five major academic databases and indexes: Google Scholar, IEEE, Springer, ACM Digital Library, and Scopus. The search focused on peer-reviewed articles published between 2020 and 2025, using the "Publish or Perish 8" tool and manual searches in IEEE, Scopus, and the ACM library. The following keyword groups and Boolean operators were used: AI agent, artificial intelligence agent, conversational agent, intelligent tutoring system, chatbot, ChatGPT, GitHub Copilot as AI Agent Keywords, and programming education, teaching programming, computer science education, introductory programming course as Education Keywords. The search string used (in various logical combinations) was: ("AI agent" OR "conversational agent" OR "intelligent tutoring system" OR "ChatGPT" OR "GitHub Copilot") AND ("programming education" OR "teaching programming" OR "computer science education").

### 3.2 Inclusion and Exclusion Criteria

The selection of articles for the study was based on the following inclusion and exclusion criteria to ensure their relevance and quality. All these criteria are shown in Table 1.

Table 1: Inclusion and Exclusion Criteria for the Selected Studies

| Criterion | Eligibility | Exclusion |
|---|---|---|
| Article type in terms of refereeing | Peer-reviewed journal or conference papers | Preprints without peer review (e.g., arXiv submissions) |
| Document Type | Peer-reviewed journal or conference papers | Non-peer-reviewed publications (e.g., blog posts, editorials) |
| Language | English | Non-English |
| Research areas | Focused on AI agents used in programming education or computer science education | Papers discussing general AI in education without a specific focus on programming |
| Research Style | Provided empirical data, case studies, or evaluations | Conceptual articles with no evaluation or application |
| Timeline | Published between 2020 and 2025 | < 2020 |

## 3.3 PRISMA Flow Diagram

A PRISMA flow diagram illustrating the article selection process, including records found, screened, and included, is provided in Figure 1:

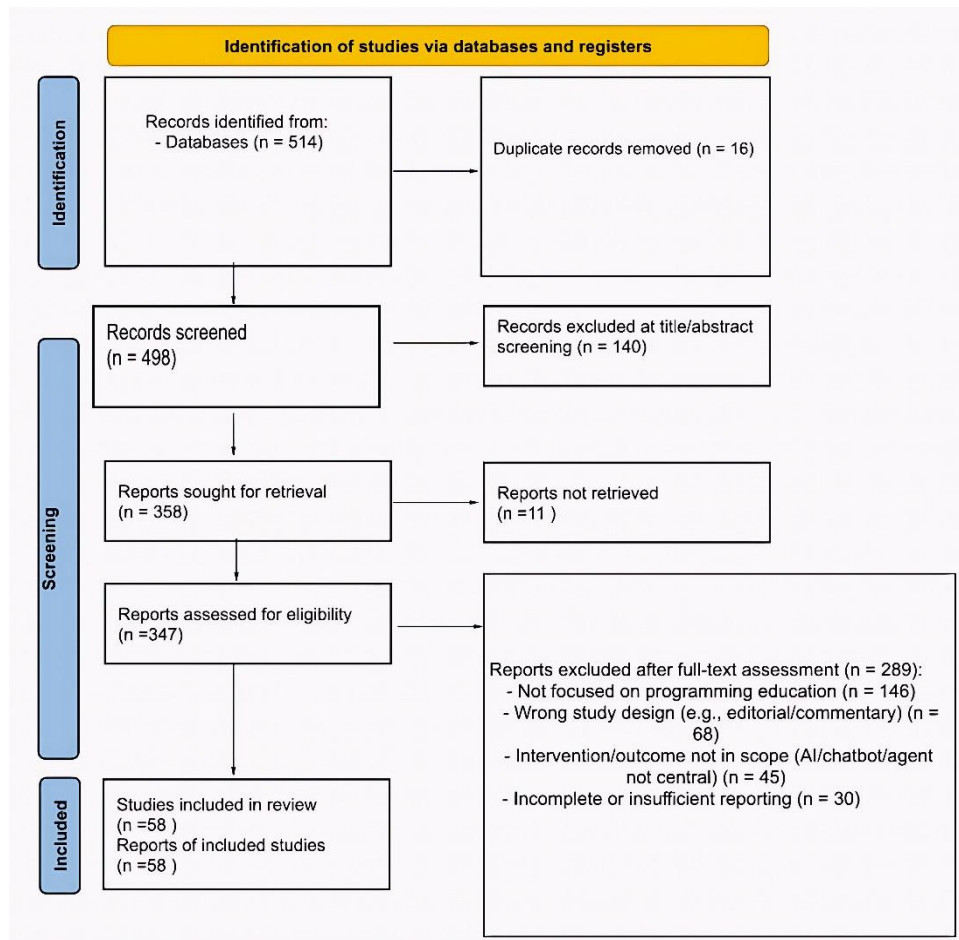

**Figure 1: PRISMA flow diagram for study selection**

## 3.4 Study Selection Process

Of the 514 articles retrieved from the databases, 16 duplicate records were removed, and 498 records were screened for title/abstract review. Of these, 140 records were discarded at the title/abstract stage, and 358 documents were requested for full text; 11 full texts were not available, resulting in 347 articles being evaluated for full text review. After a full-text review, 289 articles were removed for various reasons (the breakdown of reasons is shown in the chart), and ultimately, 58 studies were included in the final synthesis. All screening and selection processes were independently performed by the three reviewers, to ensure reliability and consistency.

## 3.5 Quality Appraisal

The quality of the studies included in this review was assessed using the Mixed Methods Appraisal Tool (MMAT 2018), which evaluates various research designs, including qualitative, quantitative, and mixed-methods studies. This tool consists of five key questions for each study. Three reviewers independently evaluated each study, considering the clarity of the study's objectives, the consistency of statistical methods, the appropriateness of the sampling strategy, the transparency of the research questions, and the presentation of findings. Out of 52 articles eligible for MMAT assessment, 18 studies met all five criteria, while 34 studies met four out of five criteria. For articles that were reviews or concept papers and could not be assessed using MMAT, the Scale for the Assessment of Narrative Review Articles (SANRA) was applied. SANRA evaluates narrative reviews based on six domains: relevance of the topic, clarity of aims, transparency of the literature search, appropriateness of referencing, soundness of scientific reasoning, and clarity of data presentation. Among the six articles evaluated with SANRA, four articles scored 10/12 and two articles scored 11/12 (Baethge et al., 2019; Hong et al., 2018). All scoring data are available in appendix 2, 3.

## 4. Data Extraction and Analysis

To facilitate a comprehensive analysis, data were extracted from each of the 58 reviewed articles to allow for a more comprehensive analysis and review. The items extracted from each article are Publication details, Type of AI agent utilized and its core functionalities, Key research findings and insights, reported benefits of AI integration in programming education and Identified challenges and limitations. The extracted data from all 58 studies were compiled into appendix 1, summarizing the focus and outcomes of each paper in a structured format.

### 4.1 Temporal Distribution of Studies (2020–2025)

To show the time trend of the 58 reviewed studies, a bar chart was created showing the number of studies published between 2022 and 2025. The results indicate that the number of studies conducted in this field increases annually, highlighting the importance of this issue. Figure 2 illustrates the temporal distribution of the reviewed studies.

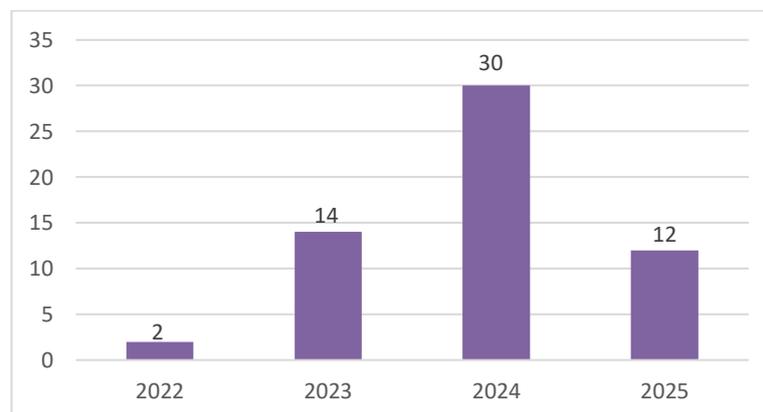

**Figure 2: Number of studies published per year (N=58)**

## 5. Results and Findings

This section synthesizes findings from 58 studies examining diverse applications of AI agents in programming education. The analysis addresses the research questions by identifying principal agent types, instructional objectives, learning outcomes, and research methodologies.

## 5.1 Classification of AI Agents in Programming Education

To address RQ1, this section presents the classification of AI agents identified across the 58 included studies. The analysis revealed three broad categories of agents. It should be noted that the category counts and percentages do not sum to the total of 58 studies, because several articles examined more than one type of AI agent and were therefore counted in multiple categories (i.e., categories overlap). Figure 3 illustrates the percentage of AI agents used in these articles. It should be noted that the statistics and percentages mentioned overlap because some articles examined multiple AI agents.

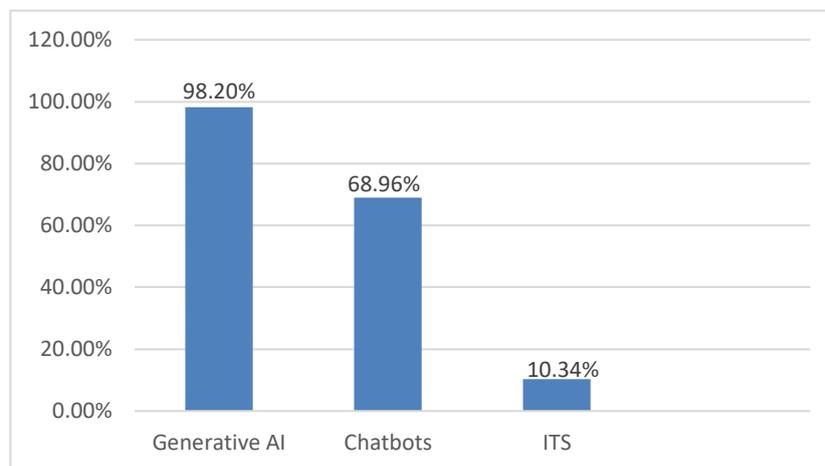

**Figure 3: Prevalence of AI agent types across reviewed studies (with overlapping)**

Based on this classification, AI agents can be grouped into the following categories:

- **Chatbots (40 studies, 68.96%):** This category includes conversational AI tools designed for real-time interaction and assistance. They are the most discussed type of AI agent in the literature.
    - **ChatGPT and other LLMs:** Chatbot systems such as ChatGPT, and Gemini are tools that process text or code inputs and produce an appropriate response. Their main functions include code generation, debugging, step-by-step explanations, and real-time feedback (Bejarano et al., 2025). In education, these systems optimize problem-solving time by providing examples and simplifying complex concepts, and their 24-hour availability enables continuous learning (Manley et al., 2024).

- **GenAI (57 studies, 98.2%):** These agents are integrated into development environments to assist with writing, suggesting, and completing code.
    - **ChatGPT:** ChatGPT, which was mentioned in 33 articles, acts as an instant tutor and debugger, providing text or code responses (Bejarano et al., 2025). ChatGPT is often used for debugging, answering questions, and generating code, increasing user interactions (Rahe & Maalej, 2025).

9— **GitHub Copilot:** This agent, which has been studied in fifteen papers, generates suggested code when prompted, commented, or incomplete code (Simaremare et al., 2024). The results show that GitHub Copilot enhances the speed and progress of problem-solving (Wermelinger, 2023). Other benefits include increasing programming speed and facilitating the implementation of basic algorithms (Nguyen & Nadi, 2022).

— **Codex and Code Llama** :Codex and Code Llama were mentioned in three studies and were examined for their effect on novice learners, showing higher scores and improved completion rates (Kazemitabaar et al., 2023). Code Llama was utilized in a study to design and evaluate an AI-assisted grading tool, which resulted in saved grading time and improved consistency (Nagakalyani et al., 2025).

— **GPT-3,3.5,4**: In 14 relevant peer-reviewed papers, various GPT models were evaluated, including GPT-3, GPT-3.5, GPT-4, and custom versions. The results showed that GPT-4 generally provides the highest accuracy and quality of responses and is more closely aligned with human behavior (Phung et al., 2023). In contrast, GPT-3 and GPT-3.5 have higher speed and scalability, but also show more errors (Alyoshyna, 2024; Deriba et al., 2024).

- **ITS and Specialized Agents (6 studies, 10.34%):** This category includes systems designed for a more structured, pedagogical approach, often with a specific learning goal.
    — **LLM-based Teachable Agents (e.g., AlgoBo):** These agents are designed to be taught by students, which can improve knowledge building and metacognitive skills (Jin et al., 2024; Chen et al., 2024). This approach emphasizes explaining concepts to the AI, which enhances the student's understanding (Chen et al., 2024).
    — **ITS (e.g., Blackbox):** A study on this ITS showed that it was most effective for intermediate and advanced students, improving their programming skills and comprehension while reducing anxiety (Fodoup Kouam, 2024)
    — **Custom GPT-based Assistants:** Several studies utilized custom-built assistants for specific purposes, such as an AI-assisted grading tool (Nagakalyani et al., 2025) or a custom VSCode plugin for CS1 students (Amoozadeh et al., 2024). These assistants offer a more controlled and focused learning experience (Akçapınar & Sidan, 2024). Liu et al. (2024) also reported that GPT-4 on CS50.ai provided 24/7 personalized instruction with 88% accuracy in course answers and 73% student satisfaction, reduced instructor workload, and improved code explanations and programming style feedback.
    — **GenAI for UI Design (Uizard):** Ho et al. (2024) focused on Uizard, an AI that reduced UI design time significantly and enabled beginners to create professional designs.

## 5.2 Instructional Objectives Supported by AI Tools

In line with RQ2, the educational purposes of AI deployment identified across the 58 reviewed studies can be organized into four major themes, as illustrated in Figure 4.

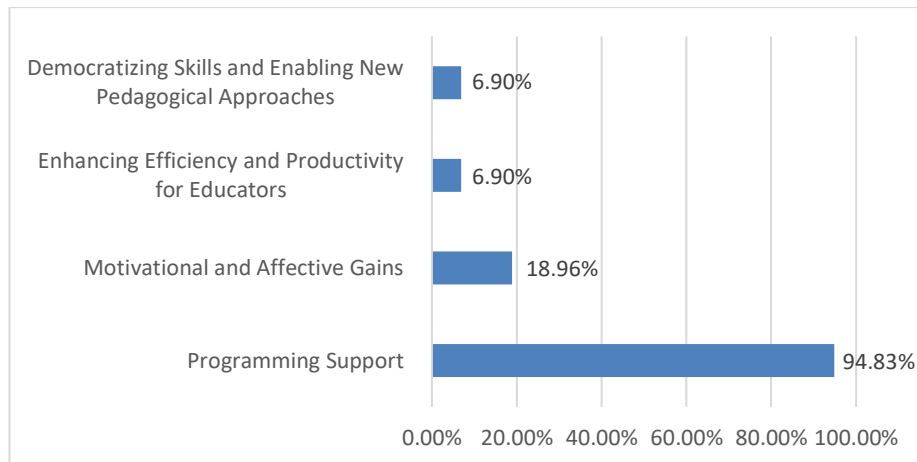

**Figure 4: Distribution of educational objectives across the reviewed studies**

- **Programming Support (55 studies, 94.83%):** As illustrated in Figure 4, programming support appeared as the most common educational objective, representing 94.83% of the reviewed studies. This objective focuses on direct assistance with coding tasks and skill development. AI tools are used to:
    — **Improve Debugging and Error Correction:** AI agents help with compiler errors and error detection (Balse et al., 2023; Pankiewicz & Baker, 2024; Deriba et al., 2024). The AI-Lab Framework improved debugging skills by 57% (Bejarano et al, 2025). Students often use AI tools like ChatGPT and LLMs for debugging (Keuning et al., 2024; Lepp & Kaimre, 2025; Manley et al., 2024; Zviel-Girshin, 2024).
    — **Enhance Code Generation and Comprehension:** Tools like GitHub Copilot accelerate coding and improve productivity (Nguyen et al., 2022; Shihab et al., 2025; Wermelinger, 2023). OpenAI Codex resulted in 1.8 times higher scores and improved completion rates for novices (Kazemitabaar et al., 2023). ChatGPT and other LLMs provide multiple solution approaches and explanations (Chang et al., 2024; Deriba et al., 2024; Kazemitabaar et al., 2023; Lepp & Kaimre, 2025). Shihab et al. (2025) reported that students spent 11% less time writing code manually when using Copilot.
    — **Facilitate Personalized Learning:** AI offers instant Q&A support and personalized feedback (Xie, 2024; Garcia, 2025; Deriba et al., 2024). This approach enables self-paced learning and provides structured guidance, with one study demonstrating a 57% improvement in debugging skills through a structured AI

approach (Bejarano et al., 2025; Denny et al., 2024; Shanto et al., 2024).

- **Motivational and Affective Gains (11 studies, 18.96%):** As shown in Figure 4, 18.96% of the studies focused on motivational and affective gains, highlighting AI's role in shaping learners' emotions and engagement.
    — **Reduce Anxiety and Frustration**: Several studies found that AI tools can reduce programming anxiety and negative emotions such as frustration (Boguslawski et al., 2025; Zviel-Girshin, 2024). AI-assisted pair programming, for instance, has been shown to minimize programming anxiety significantly (Fan et al, 2025). An ITS also had a positive impact on comprehension and reduced anxiety (Fodoup Kouam, 2024).
    — **Increase Engagement and Motivation:** AI agents, particularly in a collaborative setting, have been shown to increase student engagement, intrinsic motivation, and interest (Deriba et al., 2024; Wang et al., 2025; Xie, 2024). Boguslawski et al. (2025) found that 56% of students reported increased motivation.
    — **Improve Self-Efficacy:** AI tools can enhance students' confidence and sense of competence (Boguslawski et al., 2025). Using a GenAI chatbot with a mind mapping approach has been shown to strengthen self-efficacy (Ye et al., 2025), and some students have reported increased self-efficacy in problem-solving with AI assistance (Amoozadeh et al., 2024).

- **Enhancing Efficiency and Productivity for Educators (4 studies, 6.90%):**
As shown in Figure 4, enhancing efficiency and productivity for educators was reported in 6.90% of the reviewed studies. AI tools also serve this objective by making teaching more efficient and reducing the workload for instructors.
    — **Automate Grading and Feedback:** AI-assisted grading tools, such as one using Code Llama, saved 24% of grading time and improved consistency (Nagakalyani et al., 2025). LLMs can generate automated feedback and assessment tasks, which helps reduce the teachers' workload (Dolinsky, 2025; Garcia, 2025).
    — **Curriculum Design and Management:** The integration of AI requires and facilitates curriculum redesign (Zambach, 2025), shifting instructor roles and allowing them to create better task designs (Lau & Guo, 2023).

- **Democratizing Skills and Enabling New Pedagogical Approaches (4 studies, 6.9%):** As illustrated in Figure 4, this objective highlights how AI can create new learning opportunities and change the focus of programming education. In this context, democratization means making it easier for beginners and non-experts to access and use skills that were once restricted to specialists.
    — **Democratize Access to Skills:** In this context, "democratizing access to skills" means reducing the skill threshold required to

perform complex tasks, thereby enabling novices to engage with professional-level practices. A GenAI tool (Uizard) democratized UI design skills, helping beginners to create professional designs (Ho et al., 2024). Another tool allowed non-coders to manipulate data via English prompts, shifting the focus from coding to statistical concepts (Bien & Mukherjee, 2025).
— **Shift Pedagogical Focus:** AI requires a shift in teaching methods, moving the focus from code writing to co-design and prompt engineering (Bien & Mukherjee, 2025; Bull & Kharrufa, 2024; Denny et al., 2024; Frankford et al., 2024). Denny et al. (2024) even proposed "Prompt Problems" as a new type of programming exercise to focus on logic and computational thinking.

*5.2.1 Prompt Engineering as an Emerging Instructional Objective:*
The reviewed literature indicates that prompt engineering is essential for effective AI integration in programming education. Key findings include:
- The AI-Lab framework offers structured guidance for students to write effective prompts through precise problem descriptions, edge case identification, and debugging strategies (Manley et al., 2024).
- Prompt Problems introduce exercises where students write natural language prompts to generate correct code, shifting focus from coding to problem articulation skills (Denny et al., 2024).
- The effectiveness of AI is highly dependent on the learner's metacognitive skills, as they must know how to formulate effective prompts and critically evaluate the AI's output (Hartley et al., 2024; Amoozadeh et al., 2024).
- Effective prompting requires balancing specificity with conciseness for optimal accuracy and usability, as overly long or complex prompts can lead to a breakdown in comprehension by the AI (Sun et al., 2024).

By comparing the studies, the articles show different approaches to prompt engineering:
- **Scaffolded vs. Experiential Learning:** The AI-Lab framework uses structured exercises with clear instructions (Manley et al., 2024), while Prompt Problems rely on discovery through practice and visual aids (Amoozadeh et al., 2024; Denny et al., 2024)
- **Prompting Styles:** Structured prompts create complete responses that work well for beginners, while chat-style prompts keep students engaged but with less detail (Phung et al., 2023). Student-observed prompting methods include giving the complete problem description, a step-by-step approach, or a mixed style for specific debugging tasks (Amoozadeh et al., 2024).
- **Custom vs. General Agents:** Custom AI agents work better than general chatbots like ChatGPT because they can be built with prompts to focus only on relevant course content, preventing distractions and giving targeted support for students (Akçapınar and Sidan, 2024).

Key implications for programming education include:
- **Curriculum Integration:** Prompt engineering instruction should precede programming concept introduction to develop AI interaction skills first (Manley et al., 2024). Designing instructional strategies for prompting is crucial for future research (Sun et al., 2024).
- **Scaffolding Strategy:** Novices need structured guidance while advanced students benefit from exploratory approaches (Denny et al., 2024). A Socratic-style framework could be developed to ask students sequential, more profound questions, strengthening their problem-solving skills (Sun et al., 2024).
- **Assessment Evolution:** New rubrics must evaluate both prompt quality and code correctness, considering communication effectiveness alongside technical accuracy. Instructors can also leverage the ability of custom agents to log and analyze student-AI interactions to design better teaching interventions (Akçapınar and Sidan, 2024).
- **Research Gaps:** Future studies should address longitudinal impact on programming competency and optimal balance between AI assistance and independent problem-solving (Phung et al., 2023). A specific challenge is students' tendency to over-reliance on AI and not verify outputs, which hinders critical thinking and skill development (Amoozadeh et al., 2024).

Evidence suggests prompt engineering represents a new computational literacy essential for AI-augmented programming environments rather than an optional skill enhancement.

**5.3 Pedagogical Benefits and Implementation Challenges**
To answer research questions RQ2 and RQ3, this section is structured in two subsections. The first subsection reviews the achievements of using AI agents in programming education and discusses the benefits of using AI agents. The second subsection describes the implementation challenges of using AI agents in programming education.

*5.3.1 Benefits of learning with AI agents*
In this subsection, the main achievements of using AI agents in the programming learning process are described according to the findings of 58 studies. Figure 5 illustrates the pedagogical benefits of using AI agents in the programming learning process identified across the reviewed studies.

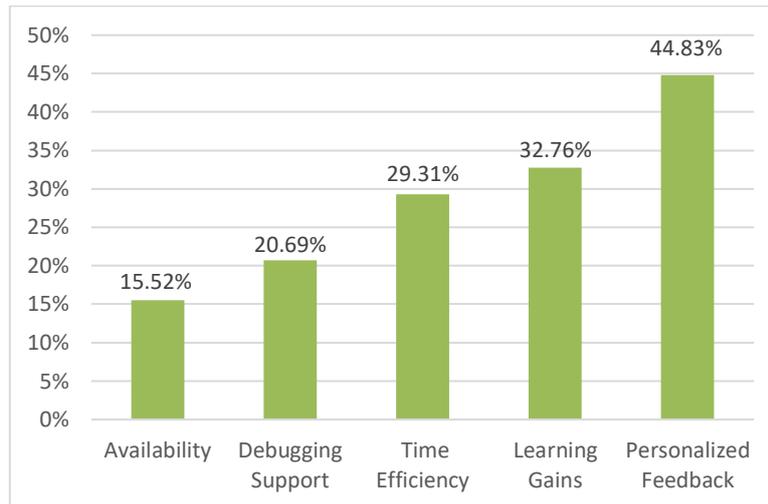

**Figure 5: Pedagogical benefits**

- **Time Efficiency (29.31%, 17/58):** A notable portion of the studies emphasized that LLMs helped reduce the time required for completing assignments, debugging, and other repetitive programming tasks. This includes time savings for both students and instructors, who saw a reduction in grading and administrative workload (Zambach, 2025; Bull & Kharrufa, 2024; Nguyen & Nadi, 2022).
- **Personalized Feedback (44.83%, 26/58):** The provision of instant, tailored feedback was a frequently reported benefit. LLMs were praised for responding adaptively to individual queries, which facilitated self-paced learning and increased student autonomy (Frankford et al., 2024; Hartley et al., 2024; Kisler et al., 2023; Shanto et al., 2024).
- **Learning Gains (32.76%, 19/58):** Several studies highlighted improvements in students' understanding of programming logic and overall comprehension of core computer science concepts. For instance, debugging skills were shown to improve by 57% with the use of a specific framework (Bejarano et al., 2025). Other benefits included higher test scores and better completion rates for novice learners (Kazemitabaar et al., 2023).
- **24/7 Availability (15.52%, 9/58):** Continuous access to AI aid was particularly appreciated in asynchronous learning settings. This enabled students to receive guidance outside of scheduled class times, whenever they needed it (Balse et al., 2023; Manley et al., 2024).
- **Debugging Support (20.69%, 12/58):** Many learners benefited from LLMs in identifying and resolving code errors. This not only enhanced efficiency but also built confidence and independence in problem-solving (Zviel-Girshin, 2024). Majority of students (75%) reported using AI for debugging purposes (Manley et al., 2024).

*5.3.2 Implementation challenges of AI agents*
The implementation challenges (based on 58 studies) will be discussed in the following section. Figure 6 presents the implementation challenges identified across the 58 reviewed studies.

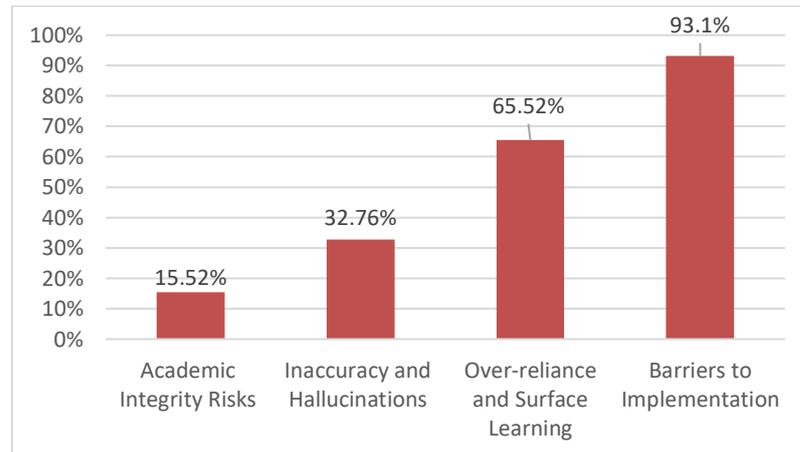

**Figure 6: Implementation challenges**

- **Over-reliance and Surface Learning (65.52%, 38/58)**: Many studies warned that students might skip deep learning by copying AI-generated code (Keuning et al., 2024; Rahe and Maalej, 2025). This leads to shallow understanding and weaker critical thinking. Concerns about overuse and dependency on AI were raised in many papers. This issue is also highlighted in papers that mention that students may not gain a deep understanding (Shanto et al., 2024), require oversight to prevent shallow learning (Kazemitabaar et al., 2023), or that the use of AI can be risky for beginners and requires basic knowledge to be effective (Nguyen et al., 2022; Wermelinger, 2023).
- **Inaccuracy and Hallucinations (32.76%, 19/58):** Several reports pointed to factually incorrect, inconsistent, or incoherent responses from LLMs. This was particularly an issue in complex programming tasks (Cámra et al., 2023; Pankiewicz and Baker, 2024) and for early models such as GPT-3.5, which showed overconfidence in incorrect answers (Kisler et al., 2023).
- **Academic Integrity Risks (15.52%, 9/58):** Concerns were raised regarding plagiarism and cheating behaviors, especially in settings with limited supervision. Some studies noted that a significant portion of students (24%) directly copied solutions from AI tools (Manley et al., 2024).
- **Barriers to Implementation (93.10%, 54/58):** The most widespread challenge involved systemic and institutional issues (Zambach, 2025). These included the need for curriculum redesign (Zambach, 2025), difficulties with assessment changes (Keuning et al., 2024; Lau and Guo, 2023) and technical limitations or high implementation costs (Xie, 2024; Chen et al., 2024). Many studies also highlighted the need for careful supervision, training, and safeguards to ensure effective integration of AI tools (Hoq et al., 2024; Lepp & Kaimre, 2025; Shanto et al., 2024)

*5.3.3 Prompt Design and Formulation Challenges*
All approaches struggle with prompt design for proper response depth—ineffective (or suboptimal) prompts tend to yield solutions that are either too simplistic or excessively complex (Rahe and Maalej, 2025). A key challenge is the difficulty in formulating logical and straightforward questions required to elicit the desired feedback from the AI (Sun et al., 2024).

Alternatively, comprehensive analysis and a more detailed review of each of the 58 studies, detailed data—including information such as the authors, publication year, the type of AI Agent employed, the target learner group, and a summary of the reported benefits and challenges are presented in full in appendix 1 .

## 6. Discussion

This systematic review examines key trends in the application of AI agents in programming education, highlighting transformative opportunities and related pedagogical challenges. The analysis extends beyond cataloging existing tools by evaluating how the field addresses ongoing pedagogical concerns.

### 6.1 GenAI and Conversational Agents: Implications for Programming Education

GenAI appeared in 98.2% of the reviewed studies, demonstrating its common use in programming education research. This shows that generative technology is no longer an accessory but has become a fundamental platform on which programming education tools are designed and implemented. In this context, chatbots remain one of the primary ways students interact with AI (68.96%). This popularity can be seen as a response to the long-standing challenges of depersonalized feedback and student frustration in introductory programming courses. The 24/7 availability and immediate, interactive support provided by tools like ChatGPT directly address the need for scalable, personalized guidance, a problem that traditional teaching methods have struggled to solve.

### 6.2 Temporal Evolution of AI in Education

The temporal analysis of the literature reveals that research in this area has experienced significant growth in a relatively brief period. The initial wave of studies in 2022–2023 was exploratory, focusing on the immediate capabilities of emerging tools such as ChatGPT and GitHub Copilot. However, research in 2024–2025 has moved towards integration with education and a focus on specialized aspects.

The emergence of specific tools for tasks such as UI design (Uizard) and automated grading (Code Llama), along with new educational frameworks such as Prompt Problems, suggests that the question has shifted from Can we use AI? to How can we use AI effectively and responsibly? This evolution suggests that the academic community is striving to create structured frameworks that both harness the benefits of AI and mitigate its risks. Temporal trends of AI agents in programming education are presented in Figure 7.

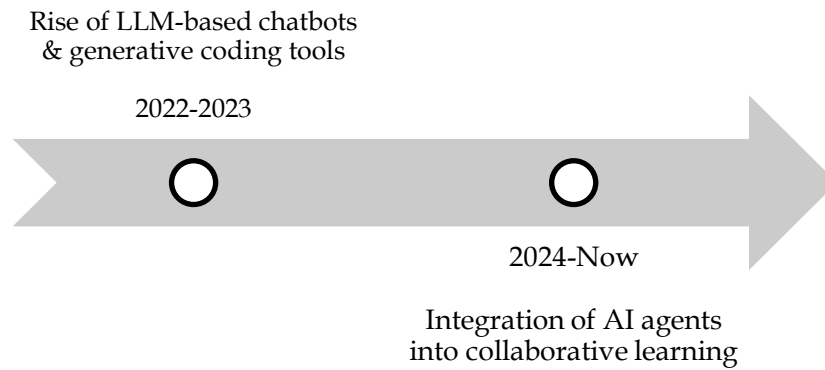

**Figure 7: Temporal trends of AI agents in programming education**

*6.2.1 Types of AI Agents Used in Programming Education*
Figure 8 summarizes the distribution of several types of AI agents used in programming education research. Some studies examined more than one type of AI agent; therefore, the values overlap.

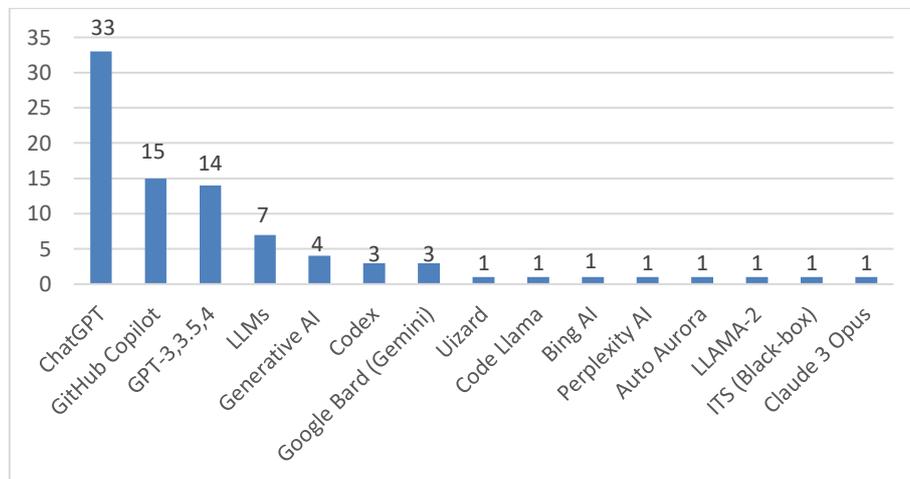

**Figure 8: Types of AI agents in programming education**

**6.3 Implications of AI Integration in Programming Education**
The integration of AI tools into programming education goes beyond usage patterns and changes the way we teach. While technical support for programming remains the most significant role, AI is also assuming new responsibilities, including enhancing student motivation and facilitating innovative teaching methods. This demonstrates that AI is not just a technical assistant, but also a cognitive and emotional framework, which can help address long-standing challenges in programming education, such as low student engagement and limited access to individualized guidance.

While fewer studies address AI's potential to enhance equitable access to programming skills, preliminary evidence suggests these tools may reduce some barriers to learning. In this context, the idea is to make it easier for beginners and non-specialists to access and use skills that were previously only available to experts. Automating routine tasks can free up instructors to focus on higher-level instructional activities while allowing novice students to engage with more complex tasks. This approach increases engagement and can inspire the design of new instructional methods.

**6.4 Pedagogical Gains and Practical Barriers of AI Use**
Studies show that using AI in programming education has its own benefits and challenges. AI can enhance teaching efficiency, offer personalized feedback, and provide ongoing support to students; however, these capabilities do not automatically foster deep learning or practical programming skills. Limitations such as inadequate infrastructure and unprepared teachers can prevent the effective use of this technology. AI can be both a supportive tool and a tool prone to error or misuse by students, and for this reason, careful management and guidance are necessary in the process of using these tools. Educators must help students interpret AI feedback correctly and evaluate the results. As a result, the successful use of AI requires access to technology, thoughtful instructional planning, and ongoing monitoring to ensure that innovation and effectiveness enhance learning.

**6.5 Prompt Engineering: An Important Skill for Programming Education**
This study demonstrates that prompt engineering, a new and essential skill, has revolutionized programming education. Students need to learn how to ask practical questions and evaluate the responses from AI. Two teaching approaches have been identified: guided methods and discovery-based methods. Both face the challenge of determining the appropriate complexity of simple questions that lead to weak answers. At the same time, overly complex ones can confuse.

Specialized tools also tend to perform better than general-purpose ones. The successful implementation of this approach requires three elements: integrating AI skills into the curriculum, developing new assessment methods, and providing step-by-step training. Excessive dependence on AI-generated solutions without proper verification remains a concern in programming education. This shift from learning programming to learning how to collaborate with AI makes prompt engineering a fundamental skill, highlighting the need for new teaching strategies.

**6.6 Synthesis and Implications: A Practical Guide for Using AI**
Based on the analysis, the following concise framework is proposed for the effective and responsible integration of AI tools in programming education:

- **Step-by-Step Problem Solving:** Utilize AI to provide hints and guidance, rather than full solutions, enabling students to think critically.
- **Prompt Skills:** Teach learners how to write clear prompts, identify errors in AI output, and evaluate results carefully.
- **Local Language and Context:** Adapt AI prompts and examples to students' language and local programming practices.

- **Ethics and Integrity:** Combine AI use with assessments, simple detection tools, and honor codes to prevent misuse.
- **Instructor Oversight:** Maintaining the instructor's active involvement is essential for explaining AI outputs, correcting misunderstandings, and guiding the learning process.

This approach helps elevate AI from mere coding assistant to a tool that supports learning, thinking, and ethical use in programming courses.

## 7. Conclusion

A comprehensive review of 58 studies on AI agents in programming education highlights key findings. The field is advancing rapidly, with notable progress in research since 2023. This growth is driven by powerful GenAI, such as ChatGPT, GitHub Copilot and GPTs. These AI tools are mainly used for programming aid, including debugging, code generation, and providing personalized feedback. Over 94% of the reviewed studies report this goal. AI agents offer significant teaching benefits but also pose notable challenges. One major benefit is the delivery of personalized feedback and 24/7 accessibility, which supports self-paced learning and reduces student frustration. However, a common challenge is the risk of over-reliance and superficial learning, identified in about 65.5% of studies. Other key concerns include errors in AI responses (32.76%) and the potential for cheating (15.52%).

The most frequently reported obstacle, appearing in 93.10% of studies, is system barriers to setup, such as curriculum adjustments and proper teacher training. These findings reveal a paradox: AI tools improve efficiency and offer quick support but may undermine deep learning and critical thinking if not used carefully. The research highlights that developing prompt skills is now a crucial competency for students. It also highlights the importance of structured guidance for effective interaction with AI agents. Future studies should investigate the long-term effects of AI on programming skills and develop teaching frameworks that strike a balance between the advantages of AI and the cultivation of independent, critical thinking skills.

## 8. Acknowledgements

During the preparation of this work the authors used Gemini in order to improve readability of the text. After using this tool, the authors reviewed and edited the content.

# Appendix

## Appendix 1: Comprehensive analysis of 58 papers on AI agents in programming education

| No. | ID | AI Agent | Key Findings | Benefits | Challenges |
|---|---|---|---|---|---|
| 1 | (Zambach, 2025) | Generative AI Chatbots | Advanced students improved by 23%. Intro students saw a 15% decline. Instructor roles shifted significantly. | Time savings for instructors. Personalized learning paths. Automated feedback. | Over-reliance on AI outputs. Requires curriculum redesign. Academic integrity concerns. |
| 2 | (Ho et al., 2024) | Generative AI (Uizard) | Reduced UI design time from 3h to 90s. Increased student satisfaction. Enabled professional designs by beginners. | Democratizes design skills. Visual feedback improves learning. Reduces instructor workload. | Limited platform support. Small sample size. Requires prompt engineering skills. |
| 3 | (Manley et al., 2024) | ChatGPT/LLMs | 75% used for debugging. 24% copied solutions directly. Positive perception shift in 68% of students. | Reduces frustration with errors. Provides alternative explanations. Available 24/7. | Solutions are often too advanced. Academic integrity issues. Requires monitoring. |
| 4 | (Keuning et al., 2024) | ChatGPT/LLMs | MSc students used AI 2.3x more. 82% reported time savings. Usage peaked during deadlines. | Faster problem-solving. Multiple solution approaches. Reduces cognitive load. | May discourage deep learning. Creates skill gaps. Requires assessment changes. |
| 5 | (Bejarano et al., 2025) | ChatGPT | Debugging skills improved by 57%. Better conceptual understanding. Balanced AI use with core skills. | Structured integration approach. Real-world problem solving. Progressive difficulty levels. | Risk of over-reliance. Needs careful implementation. Integrity concerns remain. |
| 6 | (Servin et al., 2024) | ChatGPT | Improved teamwork dynamics. Complex topics became more accessible. Enhanced problem-solving skills. | Reduced intimidation factor. Better collaboration. Practical applications. | Prompt engineering is difficult. Over-use concerns. Re- quires supervision. |
| 7 | (Xie, 2024) | ChatGPT | Increased students engagement. Improved understanding. More interactive learning. | Instant Q&A support. Personalized feedback. Always available. | Cheating risks. Technical limitations. Output verification needed. |

| No. | Year | AI Agent | Key Findings | Benefits | Challenges |
|---|---|---|---|---|---|
| 8 | (Balse et al., 2023) | GPT-4 | Effective for error detection. Worked well for basic concepts. Provided personalized support. | Reduced mentor workload. Available 24/7. Consistent feedback. | Needs human review. Limited to complex issues. May miss nuances. |
| 9 | (Chang et al., 2023) | ChatGPT, Copilot, Codex | Python most common. ChatGPT most popular. Varied usage patterns. | Improved feedback quality. Reduced workload. Multiple approaches. | Cheating risks. Dependency issues. Output quality varies. |

| No. | | AI Agent | Key Findings | Benefits | Challenges |
|---|---|---|---|---|---|
| 10 | (Rahe & Maalej, 2025) | ChatGPT | Students often copied full code. Reduced independent thinking. Varied by skill level. | Time savings. Quick solutions. Alternative approaches. | Hard to verify outputs. Creates skill gaps. Surface-level understanding. |
| 11 | (Denny et al., 2024) | GPT-4 (CodeHelp) | Students preferred guided learning. Step-by-step help is valued. Quick responses are needed. | Structured guidance. Immediate help. Consistent quality. | Dependency risk. Needs safeguards. Limited creativity. |
| 12 | (Pankiewicz & Baker, 2024) | GPT-4 | Helped with simple errors. Reduced confusion. Improved performance. | Time savings. Clear explanations. Available anytime. | Less effective for complex errors. Sometimes misleading. Limited context. |
| 13 | (Shanto et al., 2024) | ChatGPT | Comparable to traditional methods. Effective for basics. Good retention. | 24/7 access. Personalized pace. Immediate feedback. | Struggles with advanced topics. Requires oversight. Limited depth. |
| 14 | (Alpizar-Chacon & Keuning, 2025) | ChatGPT, Copilot | Increased productivity. Better learning outcomes. More experimentation. | Learning support. Time savings. Creative ideas. | Over-reliance. Incorrect outputs. Verification needed. |
| 15 | (Nagakalyani et al., 2025) | Code Llama | 24% grading time saved. Improved consistency. Better student feedback. | More accurate scoring. Detailed comments. Faster turnaround. | Long processing times. Needs rubric tuning. Limited flexibility. |
| 16 | (Zviel-Girshin, 2024) | ChatGPT | Growing student satisfaction. Improved debugging. Better commenting. | Helpful explanations. Automated suggestions. Time savings. | Over-reliance. Cheating concerns. Surface learning. |

| No. | Year | AI Agent | Key Findings | Benefits | Challenges |
|---|---|---|---|---|---|
| 17 | (Denny et al., 2024) | ChatGPT | Positive student reception. Improved thinking skills. New learning approach. | Focuses on logic. Computational thinking. Creative solutions. | Dependency concerns. Career anxiety. Needs integration. |
| 18 | (Kazemitabaar et al., 2023) | OpenAI Codex | 1.8x higher scores. Better completion rates. Improved retention. | Reduced cognitive load. Faster progress. Clear examples. | Better for prior knowledge. Limited transfer. Needs oversight. |
| 19 | (Nguyen & Nadi, 2022) | GitHub Copilot | Good for basic algorithms. Variable quality. Professional utility. | Fast code generation. Helpful suggestions. Time savings. | Buggy code. Lacks understanding. Risky for beginners. |
| 20 | (Wermelinger, 2023) | GitHub Copilot | Changed teaching methods. Altered student approach. New challenges. | Faster coding. Focus on debugging. Higher productivity. | Incorrect suggestions. Basic knowledge is needed. Verification required. |
| 21 | (Lau & Guo, 2023) | ChatGPT, Copilot | Instructors divided. Various adaptation strategies. Ongoing transition. | Reduced teaching load. Better task design. New approaches. | Cheating concerns. Dependency issues. Assessment challenges. |

| No. | Year | AI Agent | Key Findings | Benefits | Challenges |
|---|---|---|---|---|---|
| 22 | (Câmara et al., 2023) | ChatGPT | Weak UML modeling. Basic competence only. Limited effectiveness. | Good for small models. Quick drafts. Basic constraints. | Semantic errors. Large model limits. Needs improvement. |
| 23 | (Pesovski et al., 2024) | GPT-4 | Improved interaction quality. Better engagement. Varied responses. | Multiple personas. Dynamic content. Personalized approach. | Small sample. API costs. Preference for traditional. |
| 24 | (Wang et al., 2025) | LLM-based AI Agent | AI-CL group outperformed CSCL in achievement, self-efficacy, interest, and reduced mental effort. | Improved learning outcomes, personalized support, and real-time feedback. | Risk of over-reliance, limited context window in LLMs. |

| No. | Year | AI Agent | Key Findings | Benefits | Challenges |
|---|---|---|---|---|---|
| 25 | (Jin et al., 2024) | LLM-based Teachable Agent (AlgoBo) | "Why/How" questions improve learning Mode-shifting is effective | Boosts knowledge building (effect size=0.71), Lowers development cost, Supports metacognition | LLM knowledge control Un-natural dialogue flows |
| 26 | (Becker et al., 2023) | ChatGPT, GitHub Copilot | Generative AI works best as educational supplement requiring human oversight | Code generation, syntax error reduction, concept explanation, 24/7 Availability | Student over-reliance, ethical issues, output bias |
| 27 | (Dolinsky, 2025) | GenAI | GenAI enhances formative assessment, provides feedback, and aids in error correction. | Personalized feedback, reduced teacher workload | Ensuring accuracy and educational quality of AI responses |
| 28 | (Phung et al., 2023) | GPT-4, ChatGPT | GPT-4 outperforms ChatGPT and approaches human tutors in program repair, hint generation, etc. | Personalized feedback for students. Timesaving for educators (automated grading/task generation). Scalability (handles diverse student submissions). | Over-reliance on AI may hinder deep learning. Ethical concerns (e.g., plagiarism, bias in feedback). Limited interpretability of AI decisions. |
| 29 | (Garcia, 2025) | ChatGPT | ChatGPT supports personalized tutoring, code generation, and immediate feed-back. Students use it for knowledge reinforcement and debugging. Teachers leverage it for creating instructional materials and grading assistance. | 24/7 accessibility for students. Reduces teachers' workload (e.g., automated feedback). Democratizes programming education (e.g., for remote learners). | Academic dishonesty (e.g., plagiarism). Overreliance impairs critical thinking. Technical limitations (e.g., biased/inaccurate code). |
| 30 | (Logacheva et al., 2024) | GPT-4 | 96.5% clear exercises .64% shallow personalization. Positive student feedback | High exercise quality. Increased students' engagement. Unlimited practice material | Shallow personalization in most exercises. Difficulty level mismatches |
| 31 | (Boguslawski et al., 2025) | ChatGPT, Copilot | 77% of students use LLMs. 56% reported increased motivation. AI must be combined with human support | increased autonomy, and a sense of competence. Reduced negative emotions (e.g., insecurity, frustration). Faster progress in complex projects, | Cannot replace social support. Risk of over-reliance on AI. Need for critical evaluation of AI outputs |

| No. | Year | AI Agent | Key Findings | Benefits | Challenges |
|---|---|---|---|---|---|
| 32 | (Stone, 2024) | Generative AI (ChatGPT, Copilot, LLMs) | Need for human-centered research; lack of pedagogical frameworks; limited student perspective focus; recommendation for participatory, long-term studies | Accelerates learning progress; improves code explanation quality; aids debugging; reduces frustration | Over-reliance; hallucinated/incorrect outputs; plagiarism risks; negative effects on novices |
| 33 | (Lepp & Kaimre, 2025) | ChatGPT, AI Chatbots | Majority use AI for debugging and code understanding; frequent use negatively correlated with performance; students appreciate AI speed but report frustrations; educators should guide structured AI use to avoid dependency and promote critical thinking | Fast responses, 24/7 availability, help debugging, code understanding, personalized explanations | AI errors, inaccurate or over-complicated answers, limited understanding of user input, risk of over-reliance |
| 34 | (Deriba et al., 2024) | ChatGPT, Codex, Copilot, Google Bard (Gemini), GPT-3/4 | Majority of studies focus on higher education; Python most used language; ChatGPT most common tool; GenAI improves motivation and personalized learning but poses challenges in assessment and dependency; gaps in K-12 and African context research | Increased motivation, interest, engagement; personalized learning; clear code explanations; error correction | Over-reliance; inaccurate grading and feedback; inconsistent answers; ethical concerns; risk of hindering independent learning |
| 35 | (Yan et al., 2024) | ChatGPT, Bing AI | Effective communication with Gen AI evolves; collaboration deepens understanding and critical thinking; individual learner differences affect success; promotes sustainable and independent learning | Enhances meta-cognitive and self-regulated learning skills; improves communication skills; immediate responses; personalized learning pace | Difficulties in complex tasks collaboration; individual differences in ability; need for effective prompt formulation |
| 36 | (Simaremare et al., 2024) | ChatGPT, GitHub Copilot, Google Gemini, Perplexity AI | Students prefer traditional student-student pairing supplemented by GenAI; GenAI useful as assistant; lack of engagement in student-GenAI pairs; need for balanced integration | Enhances understanding of difficult concepts; generates examples and code; aids debugging and refactoring | Overreliance on AI; complex/inaccurate solutions; language barriers; narrow learning horizons; prompt formulation |

| No. | Year | AI Agent | Key Findings | Benefits | Challenges |
|---|---|---|---|---|---|
| 37 | (Stone, 2023) | ChatGPT, LLMs | GAI can support Python learning, but ethical use and teacher support are needed | Enhanced engagement, improved explanations | Ethical concerns, lack of local data |
| 38 | (Alyoshyna, 2024) | GPT-3.5, Copilot, Gemini | Copilot performed best; LLMs are promising but need refinement and human oversight | Immediate, scalable, personalized feedback | Limited accuracy, contextual issues, ethical concerns, technical cost |

| No. | Year | AI Agent | Key Findings | Benefits | Challenges |
|---|---|---|---|---|---|
| 39 | (Shihab et al., 2025) | GitHub Copilot | Students made 50% more progress in problem solving, workflow shifted from read then understand then implement to prompt then response then implement | 35% faster task completion, 11% less time writing code manually, 12% less time spent on web searches | Student concerns about not understanding generated code, Over reliance on Copilot suggestions, Differences in usage patterns between high and low performers |
| 40 | (Kiesler & Schiffner, 2023) | ChatGPT-3.5 and GPT-4 | 94.4-95.8% correct solutions, 97.2% provided textual explanations Errors due to task ambiguity and library constraints | Automated feedback for novices, supports self-paced learning, Generates multiple solution approaches | Overconfidence in incorrect responses (GPT-3.5), Sensitivity to ambiguous prompts, Potential training data bias |
| 41 | (Ma et al., 2024) | ChatGPT (GPT-4 API) | 78.9% pre-course and 100% post-course positive perception, Top uses: debugging (26%), Q and A (23%), code examples (23%), Shift from viewing ChatGPT as "conversational AI" to "program-ming tutor" | Rapid debugging assistance, Just-in-time concept explanations, Code optimization sup-port | Overreliance reducing independent problem-solving, Occasional incorrect/advanced solutions, Limited integration with IDEs |
| 42 | (Chen et al., 2024) | ChatGPT (GPT-4) | Natural dialogue enhances knowledge construction Teaching process improves SRL strategies, Readability benefits from explaining to AI, Need for controlled "mistakes" in AI responses | Improved knowledge gains ($\eta^2$ = 0.74) Better code clarity ($F$ = 7.39) & readability, Enhanced self-efficacy ($\eta^2$ = 0.75) & cognitive strategies ($\eta^2$ = 0.61) | No improvement in error-correction skills, ChatGPT's correctness reduced debug-ging practice, High implementation cost for customized prompts |
| No. | Year | AI Agent | Key Findings | Benefits | Challenges |
| 43 | (Penney et al., 2023) | LLM-based Conversational Agent (e.g., ChatGPT) | Instructors prioritize: Scaffolded problem-solving (not direct solutions), Credibility and customization Integration with coding environments | Reduces student isolation anxiety, provides real-time feedback Frees instructor time, Enhances engagement | Risk of over-reliance, Potential for incorrect teaching, Deterioration of student-instructor rapport Bias in LLM outputs |
| 44 | (Takerngsaksiri et al., 2024) | AutoAurora (StarCoder-based) | Students want explainable, personalized code completion | Increased productivity, Syntax assistance Co ding tutor | Over-reliance, Assessment challenges Code quality is-sues |
| 45 | (Bull & Kharrufa, 2024) | GitHub Copilot, ChatGPT | GAI changes programming focus from writing to co-design, requires teaching prompt engineering/code verification, pedagogical recommendations: Scaffolding and fading techniques, Contextualized assessments Fundamentals-first approach | Productivity boost for repetitive tasks, System design brainstorming, Junior developer support Context aware suggestions | Over-reliance risks, Copyright concerns, propagation, Convincing but incorrect code, High computational costs |

| No. | Year | AI Agent | Key Findings | Benefits | Challenges |
|---|---|---|---|---|---|
| 46 | (Bien & Mukherjee, 2025) | GitHub Copilot | Enables coding via English prompts without syntax learning, Shifts focus to statistical concepts + prompt engineering, Pedagogical recommendations: Teach prompt specificity, Debugging AI outputs, Use for non-majors only | Empowers non-coders to manipulate data, Reproducible workflows (R/R Markdown), Language-agnostic skills, Real-world AI tool alignment | Output randomness, Context window limitations, Variable name confusion, Over-reliance risks, requires precise prompting |
| 47 | (Frankford et al., 2024) | GPT-3.5-Turbo | LLMs shift programming education focus from code writing to co-design with AI, Requires teaching prompt engineering and code verification, Pedagogical recommendations: Scaffolding techniques, Contextualized assessments, Fundamentals-first approach | Real-time personalized feedback, Scalability for large classes, Debugging assistance, Adaptive learning support | Generic responses, over-reliance risks, API dependency, Hallucinations in feedback, Lack of interactivity, Token limits |
| 48 | (Hoq et al., 2024) | ChatGPT | ML models can detect AI-generated code with 97% accuracy. ChatGPT code is more concise and expert-like compared to student code. | High detection accuracy (up to 97%), Identifies patterns in ChatGPT-generated code, Potential for formative feedback systems | Limited to simple programming problems, requires ML expertise to implement, may not work as well for advanced courses |

| No. | Year | AI Agent | Key Findings | Benefits | Challenges |
|---|---|---|---|---|---|
| 49 | (Oli et al., 2023) | ChatGPT-3.5/4, GPT-4 Turbo, LLaMA-2 | GPT-4 with CoT achieves 0.81 correlation with human scores, the finetuned allmpnet model performs best (0.81 correlation), 0-1 scoring scale outperforms 1-5 scale | LLMs perform comparably to fine-tuned encoder models, Chain-of-thought (CoT) prompting improves accuracy, Scales to open-ended responses | Sensitive to prompt design, Struggles with numerical reasoning tasks, LLaMA-2 underperforms |
| 50 | (Liu et al., 2024) | GPT-4 (CS50.ai) | 88% accuracy in curricular answers, 50K+ users, 1.8M+ queries. Positive student feedback (73% satisfaction) | 24/7 personalized tutoring. Improved code explanations and style feedback. Reduced instructor workload | Occasional AI hallucinations. Prompt injection risks. Usage throttling needed for cost control |
| 51 | (Ma et al., 2024) | ChatGPT (GPT-4) | 4 usage patterns identified: Conceptual Learners, Code Verifiers, Practical Coders, AI-Reliant Coders. AI-Reliant Coders had higher assignment scores but lower final grades. | Quick error correction, Clear explanations, Personalized learning | Over-reliance, Incorrect responses, Lack of deep understanding |
| 52 | (Ye et al., 2025) | GenAI Chatbot (GPT-4) | Progressive mind maps GenAI significantly improved programming performance and computational thinking. Self-constructed mind maps also outperformed traditional methods. | Real-time feedback enhances self-efficacy. Mind maps structure complex concepts, improving problem-solving. | Over-reliance on chatbots may weaken independent thinking. Requires careful instructional design to avoid superficial learning. |

| No. | Year | AI Agent | Key Findings | Benefits | Challenges |
|---|---|---|---|---|---|
| 53 | (Fodouop Kouam, 2024) | ITS ("Blackbox") | Advanced students showed the most improvement in programming skills. Higher satisfaction with ITS interface among intermediate/advanced students. Positive impact on comprehension and reduced anxiety. | Personalized feedback enhances learning. Adaptive strategies improve engagement. Supports heterogeneous classrooms. | Limited course availability for specific topics. Restrictions on programming languages tools. Self-report bias in data. |
| 54 | (Hartley et al., 2024) | ChatGPT (GPT-4 with Copilot) | ChatGPT effectively consolidates instructional materials and provides personalized guidance. Strong support for planning (e.g., study schedules). Limited interactivity and assessment accuracy. | Scalable tutoring potential (Bloom's two-sigma benefit). Reduces cognitive load by integrating resources. Supports heterogeneous learners. | Assumes prior knowledge in assessments. Requires strong metacognitive skills from learners. Non-interactive compared to IDEs. |

| No. | Year | AI Agent | Key Findings | Benefits | Challenges |
|---|---|---|---|---|---|
| 55 | (Akçapınar & Sidan, 2024) | Custom GPT-3.5 Turbo Assistant | Score increase 26%. Large effect size $d = 1.56$. Enhanced programming performance. Immediate coding assistance. Customizable AI tool | 93% used an AI assistant 92% accepted incorrect AI info. 61% copied wrong answers directly. Only 8% rejected AI errors. Strong overall performance boost | Lack of critical thinking. High misinformation acceptance. Direct plagiarism via copy-paste. AI hallucination problems. Over-reliance on AI. Academic dishonesty risk |
| 56 | (Amoozadeh et al., 2024) | ChatGPT (via custom VSCode plugin) | 65% success rate with AI; 33% used full problem descriptions; Only 30% verified solutions; Mixed self-efficacy changes | Enhanced problem-solving capability. Immediate concept clarification. Step-by-step guidance. Increased self-efficacy for some | Over-reliance on full descriptions. Lack of verification behavior. Linear vs iterative approaches. Decreased self-efficacy for some |
| 57 | (Fan et al., 2025) | GPT-3.5 Turbo and Claude 3 Opus | AI-assisted pair programming significantly increased intrinsic motivation ($p < .001$, $d = 0.35$), reduced programming anxiety ($p < .001$), and improved performance compared to individual programming | Enhanced motivation, reduced anxiety, improved performance, 24/7 availability, consistent feedback | Cannot fully replicate human social interaction, lower social presence scores, potential over-reliance on AI |
| 58 | (Sun et al., 2024) | ChatGPT | ChatGPT increased debugging activity, feedback reading, and code interaction; improved perceptions of usefulness and ease of use; no significant performance difference | Enhanced engagement; personalized learning; improved user experience; increased motivation and confidence in programming | Over-reliance on ChatGPT. lack of integration with IDEs; limited support for complex logic and GUI; ethical concerns (bias, privacy, transparency) |

**Appendix 2: Detailed MMAT scoring for the 58 included studies**

| No. | Q1 | Q2 | Q3 | Q4 | Q5 |
|---|---|---|---|---|---|
| 1 | Yes | Yes | Yes | Yes | Yes |
| 2 | Yes | Yes | No | Yes | Yes |
| 3 | Yes | Yes | Yes | Yes | Yes |
| 4 | Yes | Yes | Yes | Yes | Yes |
| 5 | Yes | Yes | Yes | Yes | Can't Tell |
| 6 | Yes | Yes | No | Yes | Yes |
| 7 | Yes | Can't Tell | Yes | Yes | No |
| 8 | Yes | Yes | No | Yes | Yes |
| 9 | N/A | N/A | N/A | N/A | N/A |
| 10 | Yes | Yes | No | Yes | Yes |
| 11 | Yes | Yes | No | Yes | Yes |
| 12 | Yes | Can't Tell | Yes | Yes | Yes |
| 13 | Yes | Yes | No | Yes | Yes |
| 14 | Yes | Yes | No | Yes | Yes |
| 15 | Yes | Yes | Yes | Yes | Yes |
| 16 | Yes | Yes | No | Yes | Yes |
| 17 | Yes | Yes | No | Yes | Yes |
| 18 | Yes | Yes | No | Yes | Yes |
| 19 | Yes | Yes | No | Yes | Yes |
| 20 | Yes | Yes | Yes | Yes | Can't Tell |
| 21 | Yes | Yes | Yes | Yes | Yes |
| 22 | Yes | Yes | Yes | Yes | Yes |
| 23 | No | Yes | Yes | Yes | Yes |
| 24 | Yes | Yes | No | Yes | Yes |
| 25 | Yes | Yes | Yes | Yes | Yes |
| 26 | Yes | Yes | Yes | Yes | Yes |
| 27 | N/A | N/A | N/A | N/A | N/A |
| 28 | Yes | Yes | Yes | No | Yes |
| 29 | N/A | N/A | N/A | N/A | N/A |
| 30 | No | Yes | Yes | Yes | Yes |
| 31 | Yes | Yes | Yes | Yes | Yes |
| 32 | N/A | N/A | N/A | N/A | N/A |
| 33 | Yes | Yes | Yes | Yes | Yes |
| 34 | N/A | N/A | N/A | N/A | N/A |
| 35 | Yes | Yes | Yes | Yes | Yes |
| 36 | Yes | Yes | Yes | Yes | Yes |
| 37 | N/A | N/A | N/A | N/A | N/A |
| 38 | Yes | Yes | Yes | Yes | Yes |
| 39 | Yes | Yes | Yes | Yes | Yes |
| 40 | Yes | Yes | Yes | Yes | Yes |
| 41 | Yes | Yes | Yes | Yes | Yes |
| 42 | Can't Tell | Yes | Yes | Yes | Yes |
| 43 | Yes | Yes | Yes | Yes | Yes |
| 44 | Yes | Yes | Yes | Yes | Can't Tell |
| 45 | Yes | Yes | Yes | Yes | Can't Tell |
| 46 | Yes | Can't Tell | Yes | Yes | Yes |
| 47 | Yes | Yes | Yes | Yes | Can't Tell |
| 48 | Yes | Yes | Yes | Yes | Can't Tell |
| 49 | Yes | Yes | Yes | Yes | Can't Tell |

| 50 | Yes | Yes | Yes | Yes | Can't Tell |
| 51 | Yes | Yes | Yes | Yes | Can't Tell |
| 52 | Yes | Yes | Yes | Yes | Can't Tell |
| 53 | Yes | Yes | Yes | Yes | Can't Tell |
| 54 | Yes | Yes | Yes | Yes | Can't Tell |
| 55 | Yes | Yes | Yes | Yes | Can't Tell |
| 56 | Yes | Yes | Yes | Yes | No |
| 57 | Yes | Yes | Yes | Yes | No |
| 58 | Yes | Yes | Yes | Yes | No |

**Appendix 3: Detailed SANRA scoring for the 6 included studies**

| No. | Q1 | Q2 | Q3 | Q4 | Q5 | Q6 | SANRA Score |
|---|---|---|---|---|---|---|---|
| 9 | 2 | 2 | 2 | 2 | 2 | 1 | 11 |
| 29 | 2 | 2 | 1 | 2 | 2 | 1 | 10 |
| 27 | 2 | 2 | 1 | 2 | 2 | 1 | 10 |
| 32 | 2 | 1 | 1 | 2 | 2 | 2 | 10 |
| 34 | 2 | 2 | 2 | 2 | 1 | 2 | 11 |
| 37 | 2 | 1 | 1 | 2 | 2 | 2 | 10 |